\documentclass[twocolumn]{aastex63}
\usepackage{xcolor}
\usepackage{gensymb}
\usepackage[nointegrals]{wasysym}
\usepackage{amsmath}
\usepackage{verbatim}

\usepackage{threeparttable}
\usepackage{bm}

\newcommand{\Kepler}{\textit{Kepler}}

\shortauthors{Kunimoto \& Bryson}
\shorttitle{A Joint Transit-RV Fit}

\begin{document}

\title{Combining Transit and RV: A Synthesized Population Model}

\correspondingauthor{Michelle Kunimoto}
\email{mkuni@mit.edu}

\author[0000-0001-9269-8060]{Michelle Kunimoto}
\affiliation{Kavli Institute for Astrophysics and Space Research, Massachusetts Institute of Technology, Cambridge, MA 02139}

\author[0000-0003-0081-1797]{Steve Bryson}
\affiliation{NASA Ames Research Center, Moffett Field, CA 94901}

\begin{abstract}
We present a framework for estimating exoplanet occurrence rates by synthesizing constraints from radial velocity and transit surveys simultaneously. We employ approximate Bayesian computation and various mass-radius (M-R) relations to explore the population models describing these surveys, both separately and in a joint fit. Using this approach, we fit a planet distribution function of the form $d^{2} N/d\log{P}d\log{M} \propto P^{\beta} M^{\alpha}$, with a break in the power law in mass at $M_{b}$, to planets orbiting FGK stars with periods $P = [25, 200]$ days and masses $M = [2, 50] M_{\oplus}$. We find that the M-R relation from \citet{Otegi2020}, which lets rocky and volatile-rich populations overlap in mass, allows us to find a model that is consistent with both types of surveys. Our joint fit gives $M_{b} = 21.6_{-3.2}^{+2.5} M_{\oplus}$ (errors reflect 68.3\% credible interval). This is nearly a factor of three higher than the break from transit-only considerations and an M-R relation without such an overlap. The corresponding planet-star mass ratio break $q_{b} \sim 7\times10^{-5}$ may be consistent with microlensing studies ($q_b \sim 6\times10^{-5} - 2\times10^{-4}$). The joint fit also requires that a fraction of $F_{\text{rocky}} = 0.63_{-0.04}^{+0.04}$ planets in the overlap region belong to the rocky population. Our results strongly suggest that future M-R relations should account for a mixture of distinct types of planets in order to describe the observed planet population.
\end{abstract}

\section{Introduction}

The discoveries of thousands of planets outside the Solar System have facilitated numerous statistical studies of exoplanet demographics. The field has progressed to the point where demographics have been characterized using multiple detection methods: namely, radial velocities \citep[RV; e.g.][]{FischerValenti2005, Cumming2008, Mayor2011}, transits \citep[e.g.][]{Howard2012, Burke2015, Fulton2017}, microlensing \citep[e.g.][]{Gould2010, Suzuki2016, Jung2019}, and direct imaging \citep[e.g.][]{NielsenClose2010, Bowler2016, Naud2017, Neilson2019, Vigan2020}. These studies provide invaluable inputs to theoretical studies and the design of future exoplanet missions, and each detection technique provides different, yet complementary constraints on the underlying planet population. For instance, RV surveys probe planet orbital period and mass, and are sensitive to low-mass planets at short periods and high-mass planets with orbits up to thousands of days. Meanwhile, the transit method is the only known method capable of exploring the population as a function of planet radius. Transit surveys such as the \Kepler\ mission \citep{Borucki2010} have revolutionized our understanding of the occurrence rates of small planets in orbital periods up to hundreds of days.

Given that no single detection method can probe all facets of the exoplanet population, synthesizing results from different surveys into a cohesive picture can place more powerful constraints on the distribution of exoplanets and their properties. However, combining and comparing results from different surveys is challenging due to unique selection effects and observational biases intrinsic to each detection method.

To address these challenges, \citet{ClantonGaudi2014} developed a forward modeling framework to simultaneously fit results from RV and microlensing for the first time. Their methodology involved generating random planet populations from an assumed planet population model, converting the planet properties into those observable by each detection method, and simulating the exoplanet yields from each survey by taking into account their detection limits. By comparing the simulated yields to the observed results for each survey simultaneously, they were able to recover a single planet population model consistent with all surveys. Their methodology was extended to include direct imaging in \citet{ClantonGaudi2016}. The end result was the most comprehensive picture of the occurrence rate of giant planets in large-separation orbits around M dwarfs to date. 

So far, only the RV, microlensing, and direct imaging methods have been combined by previonis works \citep{ClantonGaudi2014, ClantonGaudi2016, Meyer2018}. In this paper, we adopt a similar forward modeling procedure as in \citet{ClantonGaudi2014, ClantonGaudi2016}, but use an approximate Bayesian computation (ABC) method to infer our planet population model parameters, and synthesize results from the RV and transit methods for the first time. We also take advantage of comprehensive completeness contours rather than the more simple detection limits used by \citet{ClantonGaudi2016}. Synthesizing yields from transit surveys with other methods has added challenges due to fundamental differences in observable planet properties, where the transit method finds a planet's radius and other methods place constraints on a planet's mass. Our analysis will thus be sensitive to our choice of mass-radius (M-R) relation to convert between these observables. However, as we will reveal through our investigations, this sensitivity allows us to find an M-R relation that yields a single population model consistent with both RV and transit surveys, while other relations do not. Our findings will have important implications for future explorations of exoplanet M-R relations.

Our analysis also requires input surveys that have well-characterized completeness, which allows us to correct for imperfect detection efficiency, and ideally reliability, which allows us to correct for the false positive rate of a catalogue. The \Kepler\ survey is a clear choice for the transit side, as it was specifically designed to support statistical inference of exoplanet occurrence rates. Both the completeness and reliability of \Kepler\ have been thoroughly characterized \citep{Thompson2018, Bryson2020}. However, RV surveys often do not publish non-detections, making statistical analysis difficult. Here, we take advantage of the HARPS + CORALIE survey presented in \citet{Mayor2011}, which is  one of the few RV studies to provide an estimate of completeness.

This paper is organized as follows. In \S\ref{sec:ABC}, we outline the ABC methodology, which provides our inference of model parameters. In \S\ref{sec:simulator}, we describe ABC's application to finding planet population models, both by fitting RV and transit population independently, as well as simultaneously in a joint analysis. In \S\ref{sec:data}, we describe the input RV and transit surveys, and in \S\ref{sec:results}, we provide the results from applying our methodology to these surveys. In \S\ref{sec:discussion}, we discuss the implications of our results for our current understanding of the exoplanet population. Finally, in \S\ref{sec:caveats} we discuss the limitations of our current methodology and explore avenues for future improvements.

\section{Methodology}\label{sec:ABC}

\subsection{Approximate Bayesian Computation}

Bayesian inference is a popular approach of statistical inference on unknown parameters, where the goal is to estimate the probability of a model given the observed data. In other words, we want to find the posterior probability distribution $P(\bm{\theta}|D)$ of a model with parameters $\bm{\theta}$, given the data $\mathcal{D}$. From Bayes' theorem, this is

\begin{equation}
    P(\bm{\theta} | \mathcal{D}) \propto P(\mathcal{D}|\bm{\theta})P(\bm{\theta}),
\end{equation}

\noindent where $P(\mathcal{D}|\bm{\theta})$ is the likelihood function, indicating the likelihood of the data given the model, and $P(\bm{\theta})$ is the prior probability, representing initial beliefs about the model parameters before the data is taken into account.

For simple models, the likelihood function can typically be derived analytically. However, for more complex models, the likelihood may be unknown or too computationally expensive to evaluate. In these cases, Approximate Bayesian Computation (ABC) steps in as a rigorous statistical framework for likelihood-free Bayesian inference. ABC bypasses the need for a likelihood function by taking advantage of our ability to forward model the observed data under investigation, and combining that with our prior information. By simulating a large number of datasets and quantifying the degree to which they match the observed data, the distribution of model parameters providing the best matches can be determined. This distribution gives the ABC posterior, which approximates the posterior probability distribution of regular Bayesian inference.

Given the numerous complexities associated with exoplanet surveys (e.g. target selection, planet search pipelines, measurement uncertainties, selection effects, instrumental effects, false positives, star-dependent detection probabilities), the calculation of exoplanet occurrence rates is one such example where it is impractical to write down an exact likelihood function \citep{Hsu2018, Hsu2019}. \citet{Hsu2018} was the first to apply ABC in the context of exoplanet occurrence rates, and it has since been adopted in several other works in the context of both grid-based and parametric planet population models \citep{Hsu2019, He2019, He2020, KunimotoMatthews2020, KunimotoBryson2020, Bryson2020b}. 

\citet{KunimotoBryson2020} showed that a likelihood approach agrees with an ABC approach when using the well-characterized \Kepler\ DR25 catalog, which addressed many of the complexities pointed out in \citet{Hsu2018, Hsu2019}. Our present work, however, is complicated by the fact that we are attempting to recover planet population models using multiple independent surveys across different detection methods simultaneously. While \citet{ClantonGaudi2016} also used a forward modeling approach to address these challenges, they assumed the correctness of a specific likelihood function. With ABC, we do not depend on such an assumption. In particular, the RV survey we use is not sufficiently characterized to address complexities like selection effects and false positives, compromising our ability to determine a correct likelihood. 

The form of ABC used here is the Population Monte Carlo ABC (PMC-ABC) algorithm proposed by \citet{Beaumont2009}, wherein multiple generations of simulated data are created and an adaptive importance sampling scheme is used to evolve the ABC posterior. We use the \texttt{cosmoabc} Python package to implement PMC-ABC \citep{Ishida2015}. A full description of the algorithm can be found in \S2 in \citet{Ishida2015}, and we summarize it here.

\subsection{ABC Algorithm}

Using the notation of \citet{Ishida2015}, the PMC-ABC algorithm starts by drawing a set of $M$ model parameters from the prior called ``particles'', $\{\bm{\theta}^{i}\}$, with $i \in [1, M]$. $M$ is chosen to be much larger than the number of samples needed to characterize the prior. For each particle, a simulated dataset $\hat{\mathcal{D}}^{i}$ is generated, and a vector of distance functions $\bm{\rho}$ is used to find the distance between each simulated dataset and the observed data, $\bm{\rho}^{i} = \bm{\rho}(\mathcal{D}, \hat{\mathcal{D}}^{i})$. The $N$ particles giving the smallest $|\bm{\rho}^{i}|$, indicating the best agreement with the observed data, are saved. These particles constitute the zeroth ``particle system'' ($\mathcal{S}_{t=0}$), and the $75\%$ quantile of all $\bm{\rho} \in S_{t=0}$ gives the distance threshold vector for the next iteration ($\bm{\epsilon}_{t=1}$).

For subsequent iterations, a parameter vector $\bm{\theta}_{\text{try}}$ is drawn from the previous particle system using importance sampling. A dataset is simulated using $\bm{\theta}_{\text{try}}$, its distance $\bm{\rho}_{\text{try}}$ to the observed dataset is calculated, and $\bm{\theta}_{\text{try}}$ is added to the particle system $S_{t}$ if $\bm{\rho}_{\text{try}} \leq \bm{\epsilon}_{t}$. This process is repeated until $N$ particles are accepted into the particle system. Particles are then assigned weights according to Eqn. (3) of \citet{Ishida2015} to facilitate the importance sampling of the next generation, and the $75\%$ quantile of all $\bm{\rho} \in S_{t}$ gives the distance threshold vector for the next iteration ($\bm{\epsilon}_{t+1}$). 

With each iteration, $\bm{\epsilon}_{t}$ gets smaller, and it becomes harder for $\bm{\epsilon}_{t}$ to be satisfied by a given set of model parameters. This means that an increasingly large number of $\bm{\theta}_{\text{try}}$ must be drawn in order to accept $N$ particles into the particle system. \texttt{cosmoabc} considers the algorithm converged when the number of draws needed is $\gg N$.

\section{ABC Applied to Exoplanet Occurrence Rates}\label{sec:simulator}

Here, we describe the forward model we use in the ABC algorithm to simulate planet catalogues given a planet population model. Adopting elements of the Exoplanet Population Observation Simulator \cite[\texttt{epos},][]{Mulders2018}, we draw planet properties according to the population model, and remove planets based on their detection completeness in order to simulate the detected exoplanet population. We define distance functions to quantify the degree to which they match the observed population of planets, and use ABC to converge on the model that provides the closest match. Our forward model will depend on whether we are attempting to simulate the exoplanet yields of an RV survey, a transit survey, or both at once.

\subsection{Population Model}

Planet populations are commonly described by one- or two-dimensional power laws, with RV surveys describing dependence on period and mass \citep[e.g.][]{Cumming2008, Pascucci2018}, and transit surveys describing dependence on period and radius \citep[e.g.][]{Mulders2018, Bryson2020}. To support our joint fit of RV and transit data, we adopt a single period-mass model. We choose mass over radius as it is more applicable to future extensions of our methodology to microlensing (for which the planet-to-star mass ratio, $q$, is an observable) and direct imaging (which places limits on a planet's mass based on its measured luminosity).

The specific choice of population model will be informed by our domain of analysis. Here, we focus on planets with masses between 2 and 50 $M_{\oplus}$, limited by incompleteness in the RV survey at the lower end and degeneracies between mass and radius at the upper end. In the giant planet regime, a given radius may correspond to any mass across several orders of magnitude. Based on a by-eye analysis of the scatter of masses and radii near this regime \cite[e.g. Figure 3 in ][]{ChenKipping2017}, we determined that this degeneracy starts to appear around $R \sim 8 R_{\oplus}$, or $M_{p} \sim 50 M_{\oplus}$. There may also be features in the mass distribution starting around $M \sim 30 M_{\oplus}$, near which forming protoplanets begin accreting gas in a runaway manner under core accretion theory, and before which a pure gas giant mass function near $\sim 100 M_{\oplus}$ is appropriate. Our choice of $50 M_{\oplus}$ does not tread too far into this transition region, while it allows us to increase both the overall reliability and number of planets in our sample. We also consider an orbital period range between 25 and 200 days. The lower limit was chosen as a point at which planets are expected to experience minimal photoevaporation \citep[e.g.][]{OwenWu2017, KunimotoMatthews2020}, while the upper limit was chosen as a balance between high completeness and increasing the number of planets in the RV sample.

Given that our orbital period range starts at 25 days and a period break for small planets is expected at $\sim$10 days \citep{Mulders2018}, we need only fit for one period power law index. There may also be a turnover in the orbital distribution between $\sim 1 - 10$ AU \cite[e.g.][]{Meyer2018, Fernandes2019}, though given that our chosen stellar sample is FGK stars, the maximum expected orbital radii are $< 1$ AU for $P < 200$ days and a single power law assumption is still valid. Meanwhile, the RV data in \citet{Mayor2011} suggests a break in mass within our domain of analysis, as discussed in \S\ref{sec:data}. \citet{Suzuki2016} and \citet{Pascucci2018} have also pointed out a peak at $\sim$8-9 $M_{\oplus}$ in the GK-planet population after converting the radii of \Kepler\ planets into masses using mass-radius relations. Thus, our planet population model is a power law in period and broken power law in mass, of the form

\begin{equation}
\begin{split}
    f(P, M | \bm{\theta}) 
    & = \frac{d^2 N}{d\log{P}d\log{M}} = \eta g(P,M),\\ 
    g(P,M) & \propto
    \begin{cases}
    P^{\beta}(M/M_{b})^{\alpha_{1}} & \text{if }M < M_{b}, \\
    P^{\beta}(M/M_{b})^{\alpha_{2}} & \text{otherwise}, \\
    \end{cases}
\end{split} \label{eqn:powerLaw}
\end{equation}

\noindent where $\bm{\theta} = (\eta, M_{b}, \beta, \alpha_{1}, \alpha_{2})$ are the model fit parameters. $M_{b}$ is the break in mass, $\beta$ is the power law index for period, and $\alpha_{1}$ and $\alpha_{2}$ are the power law indices for mass before and after the break, respectively. We assume that these power laws are independent. When Equation (\ref{eqn:powerLaw}) is normalized so that the integral over the period and mass range of interest is
\begin{equation}
    1 = \int_{M_\mathrm{min}}^{M_\mathrm{max}} 
        \int_{P_\mathrm{min}}^{P_\mathrm{max}} 
        g(P,M) d\log{P} d\log{M},
\end{equation}
then $\eta$ is the number of planets per star over $[M_\mathrm{min},M_\mathrm{max}] \times [P_\mathrm{min},P_\mathrm{max}]$.

\subsection{RV Population Simulator}

The simulator starts by randomly drawing $n_{p} = \eta n_{\star}$ periods ($P$) and masses ($M$) from the population model, where $n_{\star}$ is the number of stars in the RV stellar sample. As adopted from \citet{Mulders2018}, this involves drawing one random number to determine the period of a planet from the cumulative distribution function of the period power law, and a second to determine the mass from the cumulative distribution function of the mass power law. The simulator then converts the masses to minimum masses ($M\sin{i}$) in order to match the units of the RV observables. We find $\sin{i}$ for each planet by drawing an inclination angle $i$ according to $\cos{i} \sim \text{Uniform}(0,1)$, assuming orbital inclinations are uniformly distributed across the sky.

With the population described by period and minimum mass, the characterization of the RV survey's completeness can be used to estimate the probability of detection for each planet, $P_{\text{det}}(P, M\sin{i})$. The completeness of the RV sample is already given as a function of period and minimum mass (see \S\ref{sec:data}), so no semi-amplitude calculations or eccentricity draws are part of our simulator.
 Each planet is marked as detected if Bernoulli($P_{\text{det}}) = 1$.  The Bernoulli distribution is a special case of the Binomial distribution, where only a single trial is conducted. The resulting detected population represents our simulated catalogue of RV exoplanets.

In order to find the set of population model parameters that provide the closest match between simulated and observed planet catalogues, ABC requires a distance function to quantify their agreement. A set of model parameters is accepted by the ABC algorithm if the distance is smaller than some threshold.

First, we determine which observed planets lie within our mass range of interest by multiplying it by the median $\sin{i}_{\text{med}} \approx 0.867$ to get a corresponding minimum mass range of interest. We only include planets in our calculation of the distance if they have minimum masses in this range, and orbital periods in our period range of interest.

We then find three separate distances by comparing the orbital period distributions ($\rho_{1}$), the minimum mass distributions ($\rho_{2}$), and the sample sizes of the catalogues ($\rho_{3}$). $\rho_{1}$ and $\rho_{2}$ are calculated using the two-sample Anderson-Darling (AD) statistic, and $\rho_{3}$ is found using

\begin{equation}\label{eqn:rho3}
    \rho_{3} = \text{max}\bigg(\text{abs}\bigg(1 - \frac{l}{l_{s}}\bigg), \text{abs}\bigg(1 - \frac{l_{s}}{l}\bigg)\bigg),
\end{equation}

\noindent where $l$ and $l_{s}$ are the number of planets in the observed and simulated catalogues, respectively \citep{Ishida2015}. The two-sample AD statistic has previously been used to quantify the distance between simulated and observed planet catalogues to infer exoplanet occurrence rates with ABC \citep{He2019, He2020, Bryson2020b}. The AD statistic has also been used in likelihood-based planet occurrence rates \citep{Pascucci2018}.

The final, overall distance is then a weighted sum of the three components,

\begin{equation}\label{eqn:rho}
    \rho = \sum_{i=1}^{3} w_{i} \rho_{i},
\end{equation}

\noindent where the weights $w_{i}$ scale each term so that the most variable distance does not dominate the fit. We follow \citet{He2020} and set the weight of the $i$-th distance equal to the inverse of the root mean square (RMS) of that component, $w_{i} = 1/\sigma_{i}$. To estimate $\sigma_{i}$, we simulate 100 catalogues assuming a population model defined by a nominal set of model parameters, and find the RMS of the distances between each of pair of catalogues.

\subsection{Transit Population Simulator}

The transit population simulator has the same initial step as the RV population simulator: $n_{p} = \eta n_{\star}$ periods and masses are drawn from the population model, where $n_{\star}$ is the number of stars in the transit stellar sample. However, we require an exoplanet M-R relation to convert the drawn masses into the radii ($R$) observable with the transit method. In this paper, we will compare our results using two different M-R relations.

We first use the $2-132 M_{\oplus}$ M-R relation from \citet{ChenKipping2017}, which has previously been used to fit \Kepler\ exoplanets to a period-mass model \citep{Pascucci2018}. Given that this relation is probabilistic, we draw a radius for each planet from a normal distribution given its mass and the radius dispersion.

We use \citet{Otegi2020} for our second M-R relation. \citet{Otegi2020} suggested that the rocky and volatile-rich populations, separated by the composition line of pure-water (or alternatively, a density cut-off of $\sim$3 g/cm$^{3}$), have significant overlap in mass and radius between $5$ and $25 M_{\oplus}$ and $2$ and $3 R_{\oplus}$. Thus, for planets in this transition region, mass alone is not sufficient to assign a planet an appropriate M-R relation.

To implement the \citet{Otegi2020} results, we introduce an additional model parameter $F_{\text{rocky}}$: the average fraction of planets with masses between $5$ and $25 M_{\oplus}$ that are rocky. The simulator randomly assigns this fraction of planets to the rocky population, and the rest to the volatile-rich population. Meanwhile, all planets below $5 M_{\oplus}$ are considered rocky and all planets above $25 M_{\oplus}$ are considered volatile-rich. The corresponding M-R relation is then used to convert the mass of each population's planets to a radius. Note that an average fraction across a specific mass range is not necessarily the best representation of the data. It may be more realistic to allow the fraction to depend on mass, for example as a decreasing linear or logistic function. In addition, the start and end masses defining the overlap are also only known approximately. However, we use this simple constant fraction to reduce the risk of over-fitting the data.

Because the rocky and volatile-rich M-R relations from \citet{Otegi2020} are deterministic (of the form $(R/R_{\oplus}) = C(M/M_{\oplus})^{E}$, where $C$ is a constant and $E$ is an exponent), we use their reported uncertainties on $C$ and $E$ to reflect dispersion around the relation. The dispersion around their rocky and volatile-rich M-R relations are of similar order to the widths of the $M < 2 M_{\oplus}$ and $2 - 132 M_{\oplus}$ probabilistic M-R relations from \citet{ChenKipping2017}. After assigning a planet to the rocky or volatile-rich populations, we draw $C$ and $E$ from a normal distribution centered on the best-fit values with the associated uncertainty, and use them to calculate its radius.

With the population described by period and radius, the characterization of the transit survey's completeness can be used to estimate the probability of detection for each planet, $P_{\text{det}}(P, R)$. As for the RV population, the detected catalogue of planets is simulated by marking each planet as detected if Bernoulli($P_{\text{det}}) = 1$.

To compare the simulated catalogue with the observed catalogue, we evaluate a distance using a similar process as for RV population. First, we determine which observed planets lie within our mass range of interest by converting it into a radius range through the M-R relation. Both \citet{ChenKipping2017} and \citet{Otegi2020} find that $R \sim 1 - 8 R_{\oplus}$ corresponds to $M = 2 - 50 M_{\oplus}$. We only include planets in our calculation of the distance if they have radii in this range, and orbital periods in our period range of interest. We then find $\rho_{1}$ and $\rho_{2}$ using the two-sample AD statistic over the period and radius distributions, respectively, and $\rho_{3}$ over the sample sizes according to Equation (\ref{eqn:rho3}). We find the overall distance $\rho$ by performing a weighted sum of the three components. 

\subsection{Combined Population Simulator}

The joint fit runs the RV and transit simulators separately, using the same population model to draw both simulated populations. The RV and transit simulators output their final distances ($\rho_{\text{rv}}$ and $\rho_{\text{tr}}$, respectively), and \textit{each} distance must meet a minimum threshold in order for the ABC algorithm to accept a set of model parameters and thus converge on a shared population model. In doing so, the algorithm is able to utilize constraints from both types of surveys simultaneously. This is an important advantage over the separate fits given that some model parameters will be better constrained by the RV survey over the transit survey, and vice-versa.

\section{Input Data}\label{sec:data}

\subsection{RV Sample}

We use the combined HARPS + CORALIE survey yields reported in \citet{Mayor2011} for our RV sample. The stellar sample consists of 822 M0 to F dwarfs, selected based on low levels of activity and low rotation rates from the larger sample of $\sim1800$ stars targeted for the planet search started by CORALIE and extended by HARPS. While we do not have a stellar catalogue for the 822-large subset used by \citet{Mayor2011}, we assume that full sample described in \citet{Mortier2013} is representative of our sample, giving a median mass of $M_{s} = 0.90 M_{\odot}$ and median metallicity of [Fe/H] $= -0.08$ dex.

In total, 155 planets were detected, with 29 planets in our domain of analysis. The planet list (giving their orbital periods $P$ and minimum masses $M\sin{i}$) and star-averaged completeness contours (giving $P_{\det}(P,M\sin{i})$) are available as part of the \texttt{epos} package \citep{Mulders2018}.

\begin{figure}[ht]
    \centering
    \includegraphics[width=\linewidth]{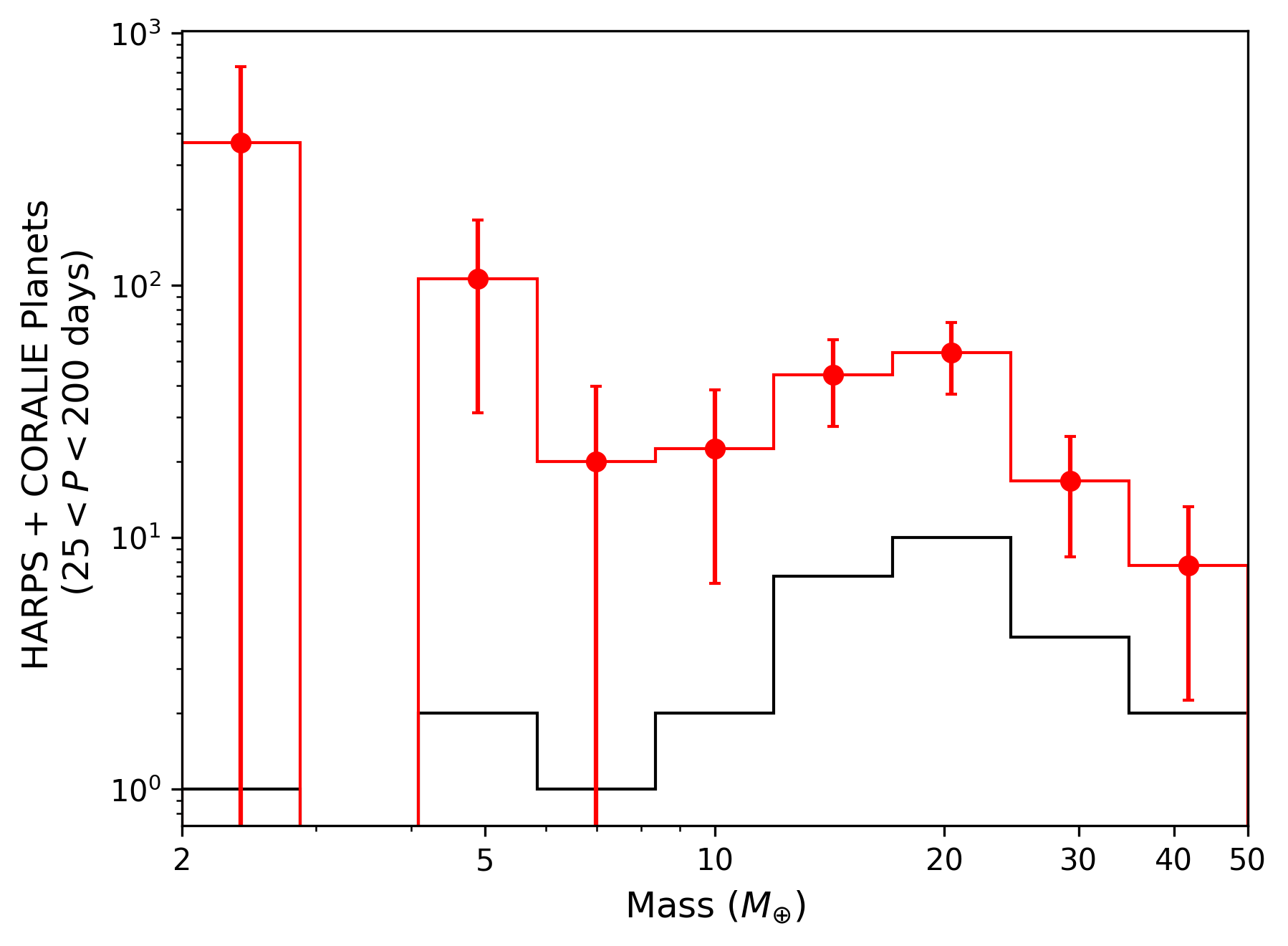}
    \caption{Histograms of planets from the HARPS + CORALIE RV surveys, comparing the observed planets (black) and the after correcting for detection bias (red). Minimum masses have been converted to an approximate mass by dividing the observed $M\sin{i}$ by $\sin{i}_{\text{med}} \approx 0.867$. }
    \label{fig:rvhist}
\end{figure}

Histograms of the masses of the 29 planets are given in Figure (\ref{fig:rvhist}), showing the distributions both before and after correction for average detection bias. At low masses, the bias-corrected histogram is dominated by uncertainty due to a small number of planets weighted by low completeness. Focusing on higher completeness past $10 M_{\oplus}$, a peak near $\sim20 M_{\oplus}$ is visible in both uncorrected and corrected histograms. This increases confidence in our ability to constrain $M_{b}$ using the RV sample.

\subsection{Transit Sample}

We use the \Kepler\ DR25 planet sample \citep{Thompson2018} associated with a clean sample of 80929 FGK stars, produced from the \citet{Berger2020} stellar properties catalogue following the procedure of \citet{Bryson2020}.  The sample has a median mass of $M_{s} = 0.97 M_{\odot}$, and median metallicity of [Fe/H] $= -0.01$ dex. 548 planets lie in the domain of analysis when using the M-R relation from \citet{ChenKipping2017} to convert the mass range into radius, compared to 544 using \citet{Otegi2020}. We use the star-averaged completeness contours from \citet{Bryson2020} to find $P_{\text{det}}(P, R)$, which takes into account the geometric probability to transit, and both detection and vetting efficiencies of the DR25 pipeline.

The contamination of the \Kepler\ catalog with astrophysical false positives and noise/systematic false alarms has been well characterized, and has been shown to have a significant impact on occurrence rates at low completeness \citep{Bryson2020}. Using the results of \citet{Bryson2020}, we find a reliability for each planet in our sample, which is an approximation of the probability that a given planet is a false positive or false alarm. We then alter our ABC distance function to accepted weighted values, and weight each observed planet according to its reliability.

\subsection{Differences Between RV and Transit Stellar Samples}

\begin{figure}
    \centering
    \includegraphics[width=\linewidth]{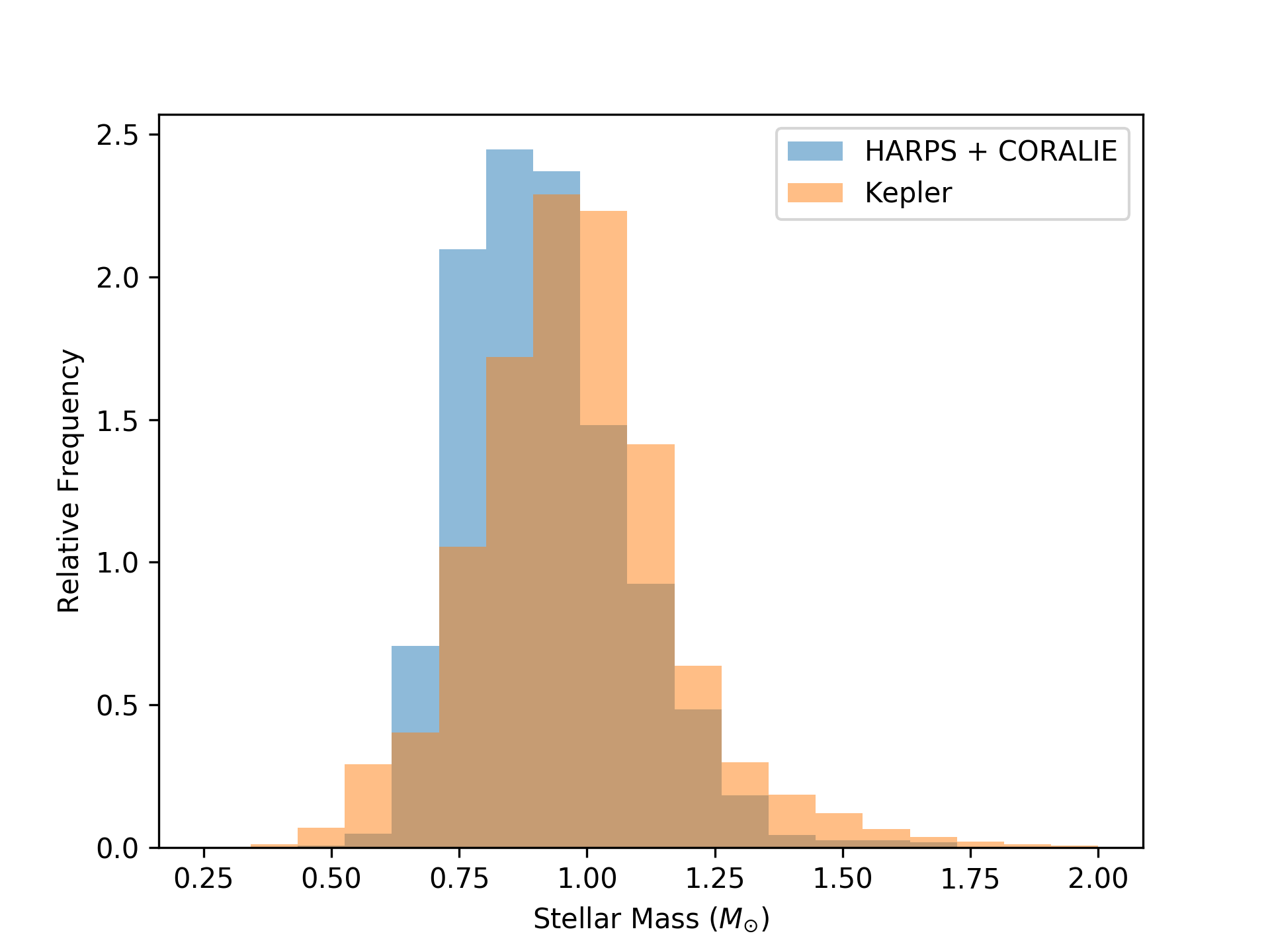}
    \includegraphics[width=\linewidth]{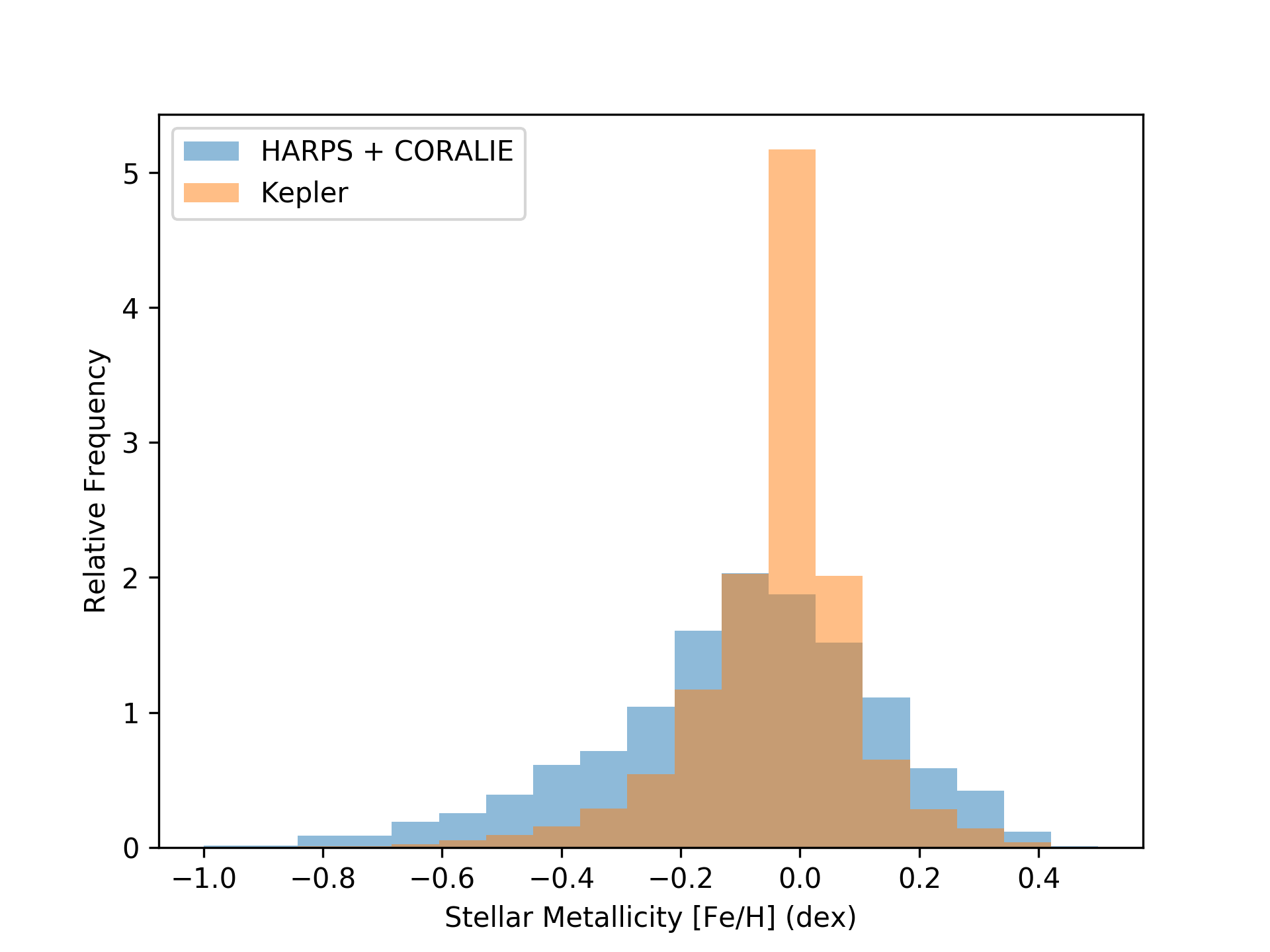}
    \caption{Comparison of the distribution of masses (top) and metallicities (bottom) for stars in the full HARPS + CORALIE sample \citep{Mortier2013} and the \Kepler\ FGK sample \citep{Berger2020}. Each histogram has been normalized, and plots do not show a small number of planets in the tails of the distributions for ease of comparison.}
    \label{fig:stellar}
\end{figure}

A concern with combining yields from independent surveys is that their stellar samples will be different, which challenges the assumption of a simultaneously consistent shared population. For example, planet occurrence rates depend on stellar properties such as mass \cite[e.g.][]{Mulders2015, KunimotoMatthews2020} and metallicity \cite[e.g.][]{Petigura2018, Narang2018}. We attempted to minimize this issue by focusing on survey yields for FGK dwarf stars only. Figure (\ref{fig:stellar}) compares the distributions of masses and metallicities between the RV and transit samples. The distributions share many of the same characteristics, with [15.9, 50, 84.1]th percentiles at [$0.76$, $0.90$, $1.07$] $M_{\oplus}$ for the RV sample and [$0.80$, $0.97$, $1.14$] $M_{\oplus}$ for the transit sample. Meanwhile, the RV sample spans a slightly wider range of metallicities with [$-0.32$, $-0.08$, -$0.12$] dex compared to the more peaked transit sample with [$-0.15$, $-0.01$, $0.05$] dex. Overall, we do not expect these differences to significantly affect our assumption of a shared mass-period power law, and proceed with the rest of the analysis with this caveat in mind.

\section{Results}\label{sec:results}

We first fit each survey separately to determine if the RV and transit populations can be described by the same population model. In all cases, we ran the ABC algorithm with \texttt{cosmoabc} five times and concatenated the posteriors to avoid the reporting of an outlier as a result. We chose wide, uniform priors for each parameter, which are given in Table (\ref{tab:results}). Table (\ref{tab:results}) also gives a summary of our results without accounting for reliability in the transit population, while Table (\ref{tab:reliability}) gives results after accounting for reliability. The central values and uncertainties are from the median and 15.9th and 84.1th percentiles of the ABC posteriors. Our results are not significantly affected by the incorporation of reliability, reflecting the fact that our period and radius range has relatively high reliability. Nevertheless, for the remainder of this paper, we will reference our reliability-incorporated results as they are expected to be a more accurate representation of the true population \citep{Bryson2020}.  Note, however, that the \citet{Mayor2011} RV sample does not have a characterized false positive rate, so we are unable to do the same exercise, and all results do not account for reliability in the RV sample.

\renewcommand{\arraystretch}{1.5}
\begin{table*}[ht]
    \centering
    \caption{ABC Fit Results}\label{tab:results}
    \begin{tabular}{c | c c c c | c}
    \hline\hline
        Parameter & RV Only & Transit Only & Transit Only & RV + Transit & Prior \\
        & & (CK17 M-R) & (O20 M-R) & (O20 M-R) & \\
         \hline
        $\eta$ & $0.5_{-0.2}^{+0.3}$ & $0.60_{-0.03}^{+0.03}$ & $0.56_{-0.03}^{+0.03}$ & $0.49_{-0.05}^{+0.05}$ & $\mathcal{U}(0,3)$\\
        $M_{b}/M_{\oplus}$ & $25.7_{-7.6}^{+8.7}$ & $8.2_{-0.9}^{+1.2}$ & $25.7_{-0.6}^{+0.8}$ & $21.5_{-3.0}^{+2.4}$ & $\mathcal{U}(2,50)$\\
        $\beta$ & $0.1_{-0.3}^{+0.3}$ & $0.1_{-0.1}^{+0.1}$ & $0.1_{-0.1}^{+0.1}$ & $0.1_{-0.1}^{+0.1}$ & $\mathcal{U}(-5,5)$\\
        $\alpha_{1}$ & $0.0_{-0.8}^{+1.7}$ & $1.3_{-0.4}^{+0.5}$ & $1.3_{-0.2}^{+0.2}$ & $1.0_{-0.4}^{+0.4}$ & $\mathcal{U}(-10,10)$\\
        $\alpha_{2}$ & $-5.0_{-3.0}^{+2.7}$ & $-2.9_{-0.7}^{+0.5}$ & $-8.5_{-1.0}^{+1.3}$ & $-6.4_{-2.2}^{+2.2}$ & $\mathcal{U}(-10,10)$\\
        $F_{\text{rocky}}$ & - & - & (0.8) & $0.62_{-0.04}^{+0.05}$ & $\mathcal{U}(0.6,1)$\\
    \end{tabular}
    \tablecomments{Central values and uncertainties are the median and 15.9th and 84.1th percentiles of the ABC posteriors. CK17 denotes \citet{ChenKipping2017}, and O20 denotes \citet{Otegi2020}.}
\end{table*}

\renewcommand{\arraystretch}{1.5}
\begin{table*}[ht]
    \centering
    \caption{ABC Fit Results (With Reliability)}\label{tab:reliability}
    \begin{tabular}{c |c c c| c}
    \hline\hline
        Parameter & Transit Only & Transit Only & RV + Transit & Prior \\
        & (CK17 M-R) & (O20 M-R) & (O20 M-R) & \\
        \hline
        $\eta$ & $0.55_{-0.03}^{+0.03}$ & $0.52_{-0.03}^{+0.03}$ & $0.47_{-0.05}^{+0.05}$ & $\mathcal{U}(0,3)$\\
        $M_{b}/M_{\oplus}$ & $8.2_{-1.0}^{+1.2}$ & $25.6_{-0.7}^{+0.7}$ & $21.6_{-3.2}^{+2.5}$ & $\mathcal{U}(2,50)$\\
        $\beta$ & $0.1_{-0.1}^{+0.1}$ & $0.1_{-0.1}^{+0.1}$ & $0.1_{-0.1}^{+0.1}$ & $\mathcal{U}(-5,5)$\\
        $\alpha_{1}$ & $1.5_{-0.5}^{+0.7}$ & $1.3_{-0.2}^{+0.2}$ & $1.1_{-0.3}^{+0.4}$ & $\mathcal{U}(-10,10)$\\
        $\alpha_{2}$ & $-3.2_{-0.9}^{+0.6}$ & $-8.8_{-0.8}^{+1.3}$ & $-6.6_{-2.1}^{+2.2}$ & $\mathcal{U}(-10,10)$\\
        $F_{\text{rocky}}$ & - & (0.8) & $0.63_{-0.04}^{+0.04}$ & $\mathcal{U}(0.6,1)$\\
    \end{tabular}
    \tablecomments{Central values and uncertainties are the median and 15.9th and 84.1th percentiles of the ABC posteriors. CK17 denotes \citet{ChenKipping2017}, and O20 denotes \citet{Otegi2020}.}
\end{table*}

\subsection{RV Fit Results}\label{sec:rv_results}

\begin{figure}[t]
    \centering
    \includegraphics[width=\linewidth]{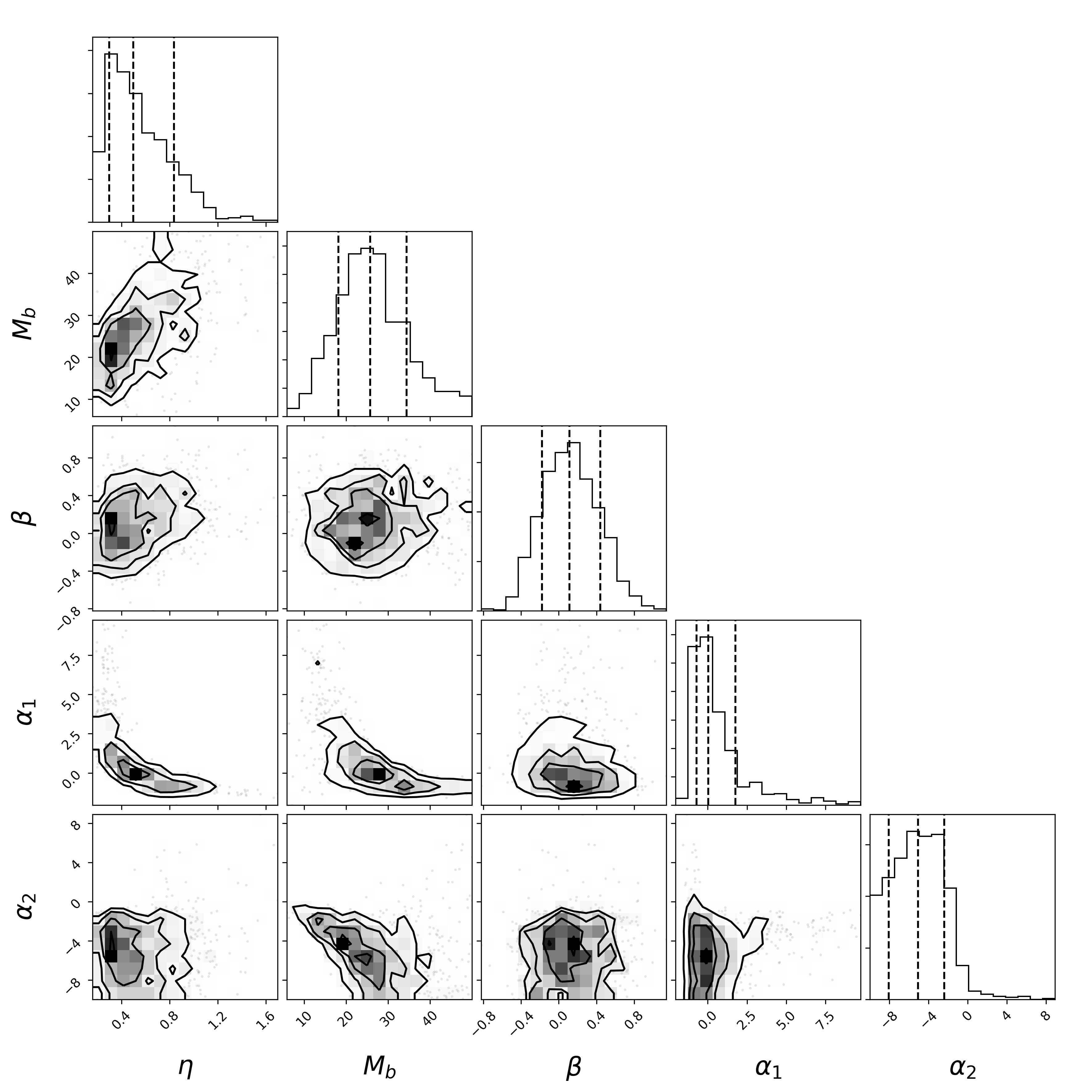}
    \caption{ABC posterior distributions for the RV-only fit. Each dotted line indicates the 15.9th, 50th, and 84.1th percentiles. Note that all histograms span a subset of our prior range (see Table (\ref{tab:results})).}
    \label{fig:rv}
\end{figure}

Our RV-only posteriors are shown in Figure (\ref{fig:rv}). Likely due to having so few planets in the domain of analysis, the posteriors are typically wide and non-Gaussian, though we see that the dataset is able to recover a mass break peaked at $25.7_{-7.6}^{+8.7} M_{\oplus}$. This is slightly higher, though consistent with the location expected from the histogram of observed planets in Figure (\ref{fig:rvhist}). We also find a drop in the mass distribution after the break with slope $\alpha_{2} = -5.0_{-3.0}^{+2.7}$, which is steeper than the mass distribution slope for giant planets, e.g. $-0.31\pm0.2$ \citep{Cumming2008}, $-0.46\pm0.06$ \citep{Fernandes2019}.

Much of our analysis depends on understanding the mass distribution from the RV dataset. Thus, before proceeding, we confirmed that a model with a mass break is indeed warranted over one with no mass break by re-fitting our RV data under an alternative no-mass-break model, of the form

\begin{equation}
\begin{split}
    f(P, M | \bm{\theta}) 
    & = \frac{d^2 N}{d\log{P}d\log{M}} = \eta g(P,M),\\ 
    g(P,M) & \propto P^{\beta}M^{\alpha}
\end{split}
\end{equation}

\noindent where $\bm{\theta} = (\eta, \beta, \alpha)$ are the new model fit parameters. We found $\eta = 0.9_{-0.3}^{+0.3}$, $\beta = 0.1_{-0.3}^{+0.3}$, and $\alpha = -0.9_{-0.3}^{+0.3}$. 
To perform a comparison between the two models, we simulated 100,000 RV populations by randomly drawing from our final ABC posteriors, first with the mass break, and then without, and found their distances to the observed RV dataset. We consider a simulation ``accepted'' if it returns a total distance smaller than some threshold. The preferred model will be the one that is more likely to recover the observed data, i.e. have more accepted simulations at small thresholds. This procedure follows that of \citet{LohmannDitlevsen2019} for model selection, which uses the ratio of acceptances as an ABC approximation of the Bayes factor, where we sample from the posterior rather than the prior.

\begin{figure}[ht]
    \centering
    \includegraphics[width=\linewidth]{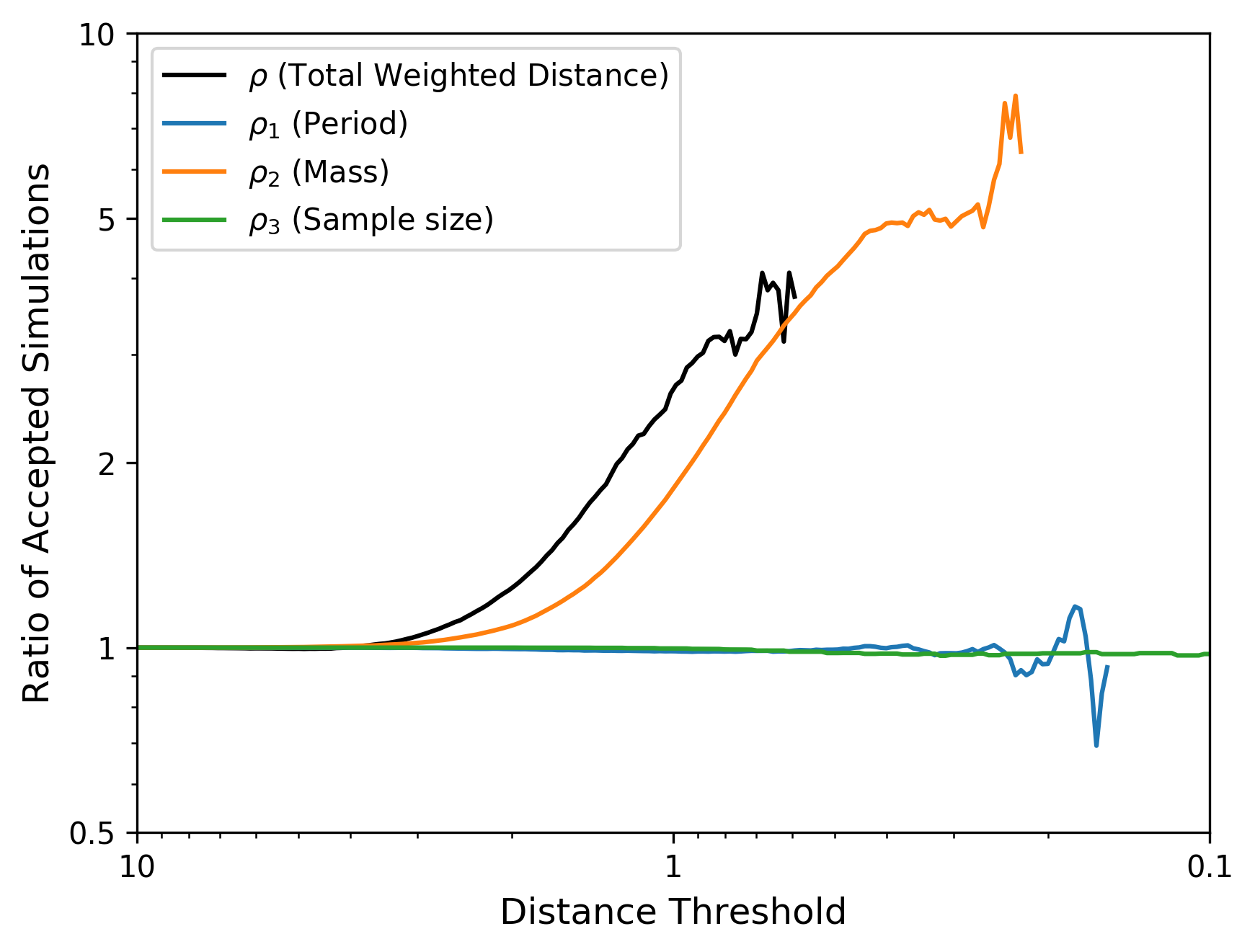}
    \caption{The number of accepted simulations with the mass break model divided by the number without the mass break model (ratio of accepted simulations), as a function of acceptance distance threshold. Shown are results when calculating the overall distance (black), as well as the  orbital period (blue), sample size (green), and mass (orange) distance components separately. Both models recover the period distribution and sample size of the observed RV dataset equally well, with roughly the same number of accepted simulations (ratio $\sim$ 1). However, the mass break component, and therefore the overall distance, show that the model with the mass break becomes increasingly favoured at small distances. Note that as the distance threshold gets smaller, the behaviour becomes noisier as it becomes exceedingly difficult for either model to satisfy such small thresholds. Only a few of the 100,000 draws satisfy the threshold for either model.}
    \label{fig:rv_comparison}
\end{figure}

In Fig. (\ref{fig:rv_comparison}), we plot the number of accepted simulations with the mass break divided by the number of accepted simulations without the mass break, and observe how it changes as a function of chosen distance threshold. We also show the results of repeating this process, but only looking at one of the three distance components (period, mass, or sample size), rather than the total weighted sum of distances. Regardless of the chosen distance threshold, the models perform equally well across period and sample size. However, the mass break model is indeed able to more frequently recover a close match to the observed mass break distribution. Note that the plot becomes more noisy at the smallest distances, as it becomes exceedingly difficult for either model to meet these thresholds.

For the remainder of our analysis, we only consider the model with the mass break.

\subsection{Transit Fit Results}

\begin{figure}[ht]
    \centering
    \includegraphics[width=\linewidth]{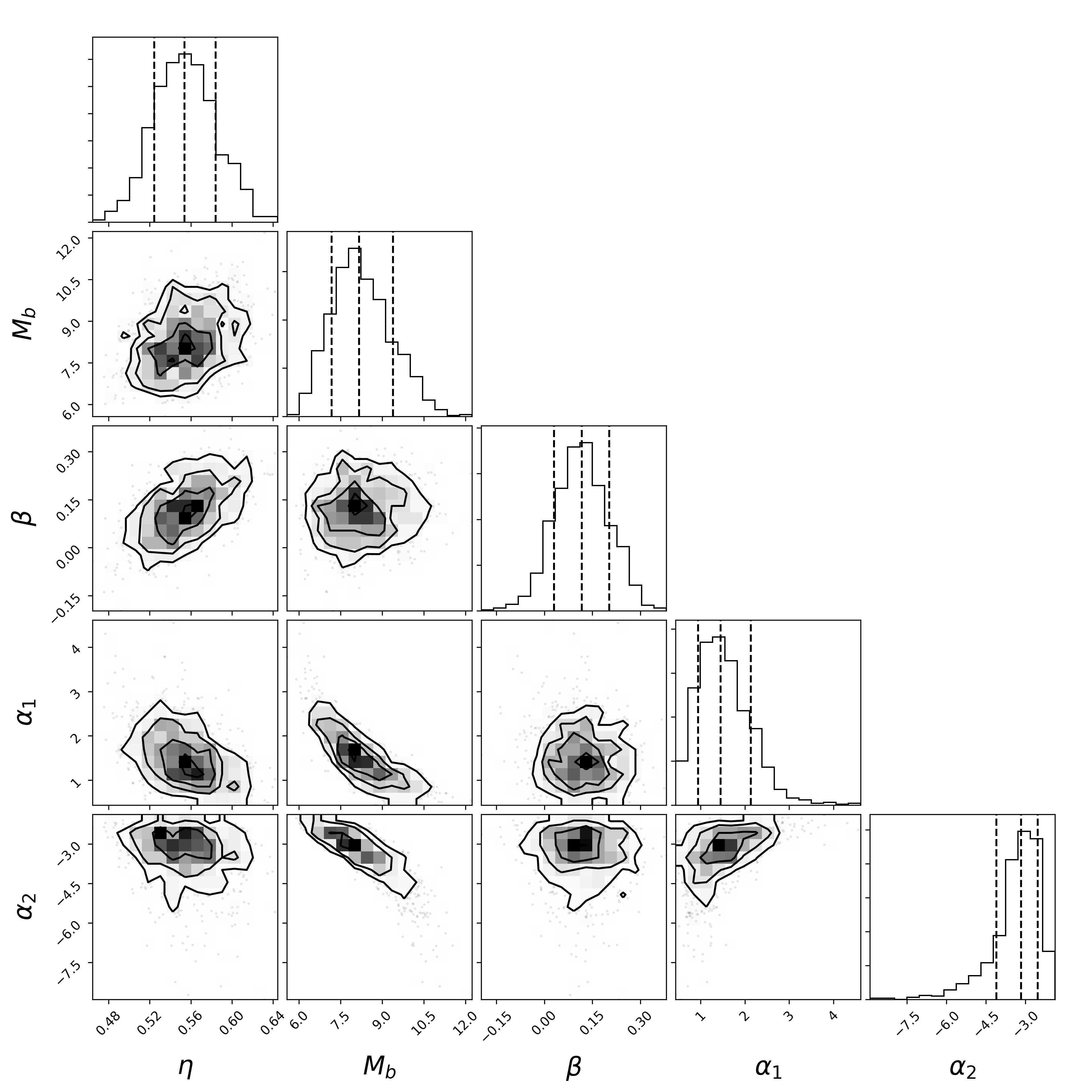}
    \caption{ABC posterior distributions for the reliability-incorporated transit-only fit using the M-R relation from \citet{ChenKipping2017}. Each dotted line indicates the 15.9th, 50th, and 84.1th percentiles. Note that all histograms span a subset of our prior range (see Table (\ref{tab:reliability})).}
    \label{fig:transit_CK}
\end{figure}

Our transit-only results, having used the \citet{ChenKipping2017} M-R relation. are shown in Figure (\ref{fig:transit_CK}). The \Kepler\ dataset is able to place much tighter constraints on all fit parameters, and encouragingly, most parameters are within 1$\sigma$ of the RV results. However, we find a significantly lower mass break at $M_{b} = 8.2_{-1.0}^{+1.2} M_{\oplus}$ (2.3$\sigma$ lower than the RV result), consistent with \citet{Pascucci2018}. This disagreement could indicate that the \Kepler\ exoplanet population is fundamentally different than the RV population. 

\begin{figure}[ht]
    \centering
    \includegraphics[width=\linewidth]{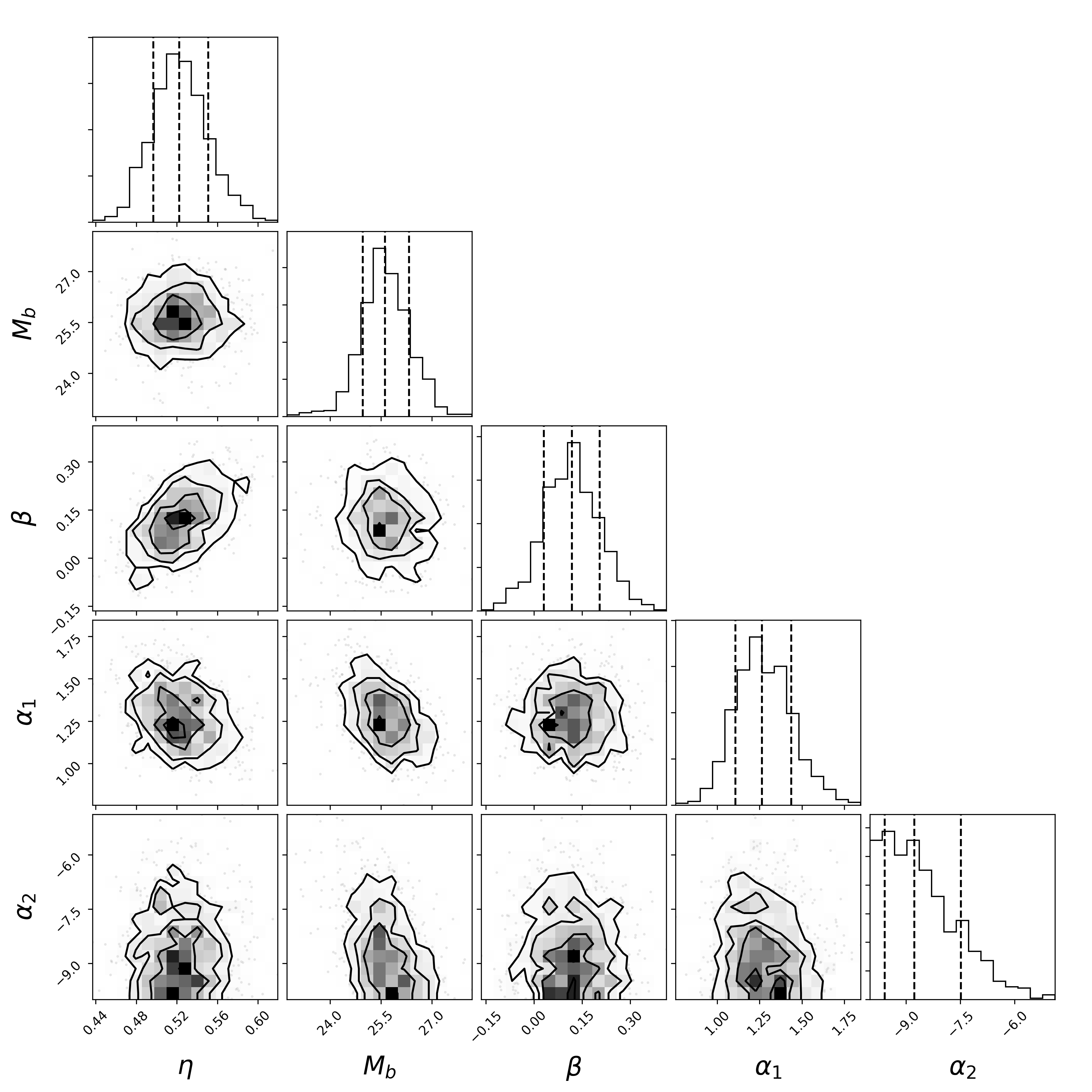}
    \caption{ABC posterior distributions for the reliability-incorporated transit-only fit using the M-R relation from \citet{Otegi2020} and setting $F_{\text{rocky}} = 0.8$ as an example. Each dotted line indicates the 15.9th, 50th, and 84.1th percentiles. Note that all histograms span a subset of our prior range (see Table (\ref{tab:reliability})).}
    \label{fig:transit}
\end{figure}

We then investigated a series of fits using the \citet{Otegi2020} M-R relation, with the goal of finding whether or not there exists an $F_{\text{rocky}}$ value that can resolve the discrepancy in $M_{b}$. Given that a rocky planet at a certain radius will have a higher mass than a volatile-rich planet at the same radius, a higher $F_{\text{rocky}}$ will have the effect of pushing the mass break to higher values. We re-ran our transit-only fits, trying $F_{\text{rocky}} \in \{0.0, 0.1, 0.2, ..., 0.9, 1.0\}$. Results for $F_{\text{rocky}} = 0.8$ as an example are given in Figure (\ref{fig:transit}).

\begin{figure}[ht]
    \centering
    \includegraphics[width=\linewidth]{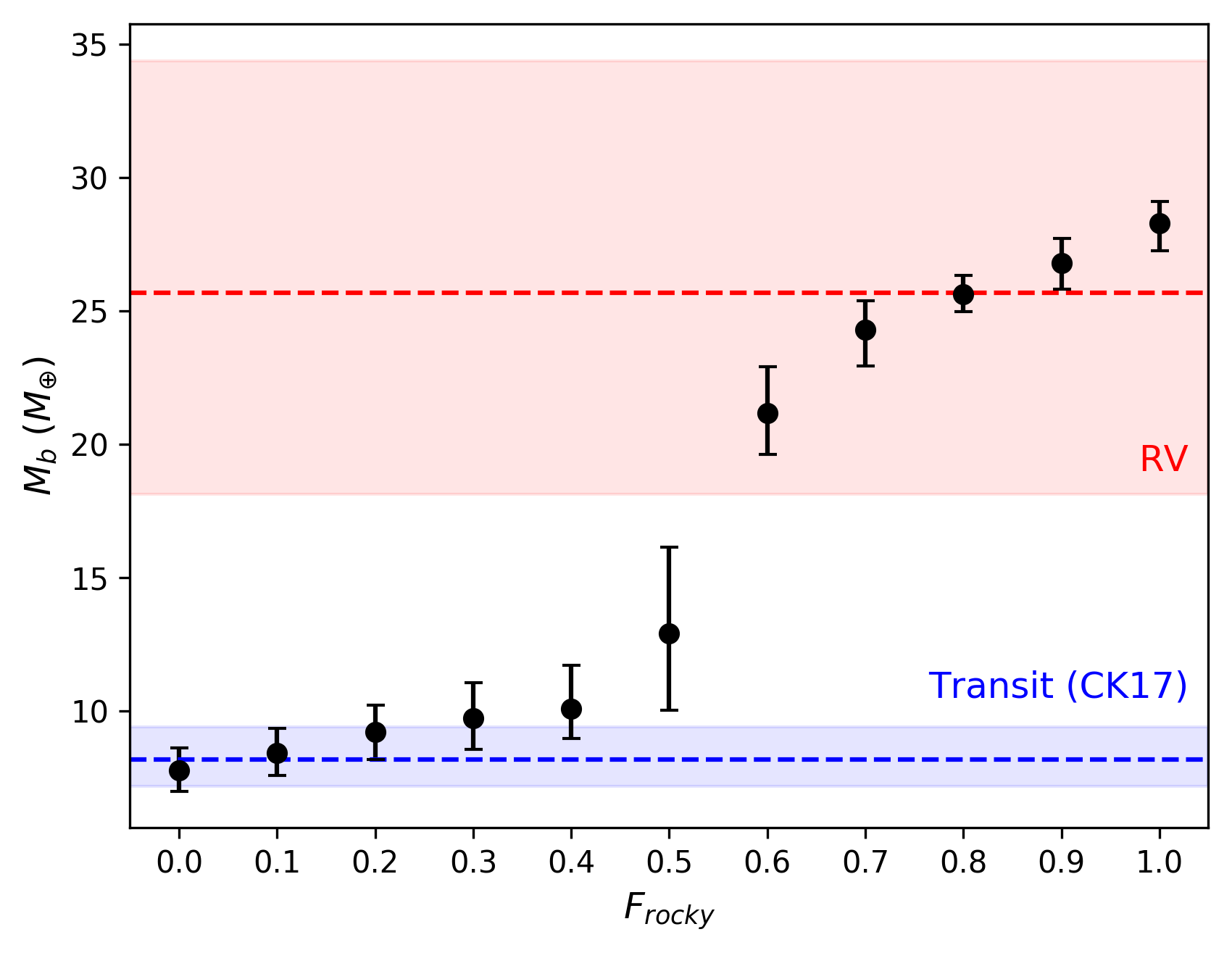}
    \caption{Resulting $M_{b}$ values from using the \citet{Otegi2020} M-R relation and setting $F_{\text{rocky}}$ fixed to various values. The blue area covers the mass break after using the \citet{ChenKipping2017} M-R relation, while the red area covers the mass break inferred from RV.}
    \label{fig:frocky}
\end{figure}

Across all $F_{\text{rocky}}$ values, the $\eta$, $\beta$, and $\alpha_{1}$ parameters are within $1\sigma$ of the \citet{ChenKipping2017} results. Meanwhile, $M_{b}$ becomes significantly larger and $\alpha_{2}$ steepens with increasing $F_{\text{rocky}}$. Figure (\ref{fig:frocky}) demonstrates how the mass break changes with $F_{\text{rocky}}$. The lowest $F_{\text{rocky}}$ values returned a mass break similar to the \citet{ChenKipping2017} result, with $F_{\text{rocky}} \in \{0.0, 0.1, 0.2, 0.3\}$ giving $M_{b}$ within $1\sigma$. However, $F_{\text{rocky}} \in \{0.6, 0.7, 0.8, 0.9, 1.0\}$ all returned mass breaks within $1\sigma$ of the RV result.

We performed a similar model comparison analysis as in \S\ref{sec:rv_results}, but this time to confirm that using the \citet{Otegi2020} M-R relation indeed allows us to better recover the observed RV mass distribution over the \citet{ChenKipping2017} M-R relation. We simulated 100,000 RV populations from the \citet{Otegi2020} and \citet{ChenKipping2017} final posteriors. We found their total distances to the observed RV data, and accepted those whose distances from the RV data were less than a given threshold.

\begin{figure}[ht]
    \centering
    \includegraphics[width=\linewidth]{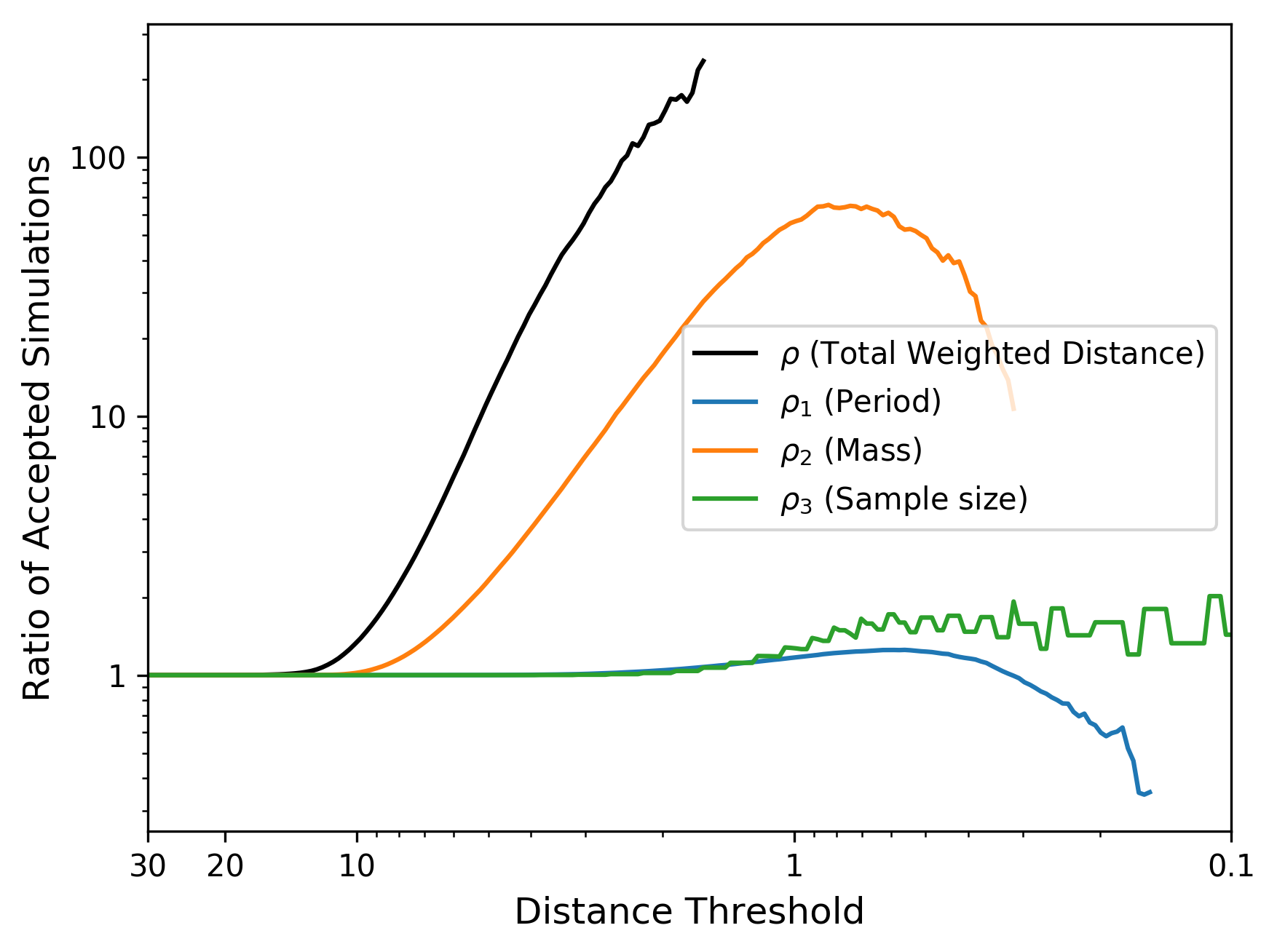}
\caption{The number of accepted simulations with the \citet{Otegi2020} M-R relation and $F_{\text{rocky}} = 0.8$ as an an example, divided by the number with the \citet{ChenKipping2017} M-R relation (ratio of accepted simulations) as a function of acceptance distance threshold. Shown are results when calculating the overall distance (black), as well as the  orbital period (blue), sample size (green), and mass (orange) distance components separately. Both models recover the period distribution and sample size of the observed RV dataset equally well, with roughly the same number of accepted simulations (ratio $\sim$ 1). However, the mass break component, and therefore the overall distance, show that the \citet{Otegi2020} model becomes increasingly favoured at small distances. Similar to Fig. \ref{fig:rv_comparison}, for distances less than $\sim 1$, it becomes more difficult for either model to be accepted. Only a few of the 100,000 draws satisfy the threshold for either model, so we do not believe the results in this regime are significant.}
    \label{fig:transit_comparison}
    \end{figure}

Figure (\ref{fig:transit_comparison}) shows the ratio of accepted simulations, dividing the results from \citet{Otegi2020} and $F_{\text{rocky}} = 0.8$ with the results from \citet{ChenKipping2017}. We also show performance across the three distance components separately. The \citet{Otegi2020} model is strongly preferred for recovery the mass distribution of the RV data. Note that as before, the behaviour of these plots becomes noisier and at the smallest distance thresholds, as it becomes exceedingly difficult for either model to satisfy such small thresholds.

\subsection{Joint Fit Results}

Having confirmed that the \Kepler\ population can plausibly be described with a mass break consistent with RV data, we now turn to our combined fit. This fit has an important advantage over the previous analyses as it can utilize constraints from both types of surveys at once. In particular, the RV side informs $M_{b}$ given that the result is independent of an M-R relation, while the transit side better constrains $\eta$, $\beta$, and $\alpha_{1}$. 

Given that our transit-only investigations indicated that any of $F_{\text{rocky}} \in \{0.6, 0.7, 0.8, 0.9, 1.0\}$ may be consistent with RV, we placed an initial uniform prior of $\sim\mathcal{U}(0.6, 1)$ on $F_{\text{rocky}}$ for the combined fit, though in subsequent iterations of the algorithm, all values are constrained between 0 and 1. 

\begin{figure}[ht]
    \centering
    \includegraphics[width=\linewidth]{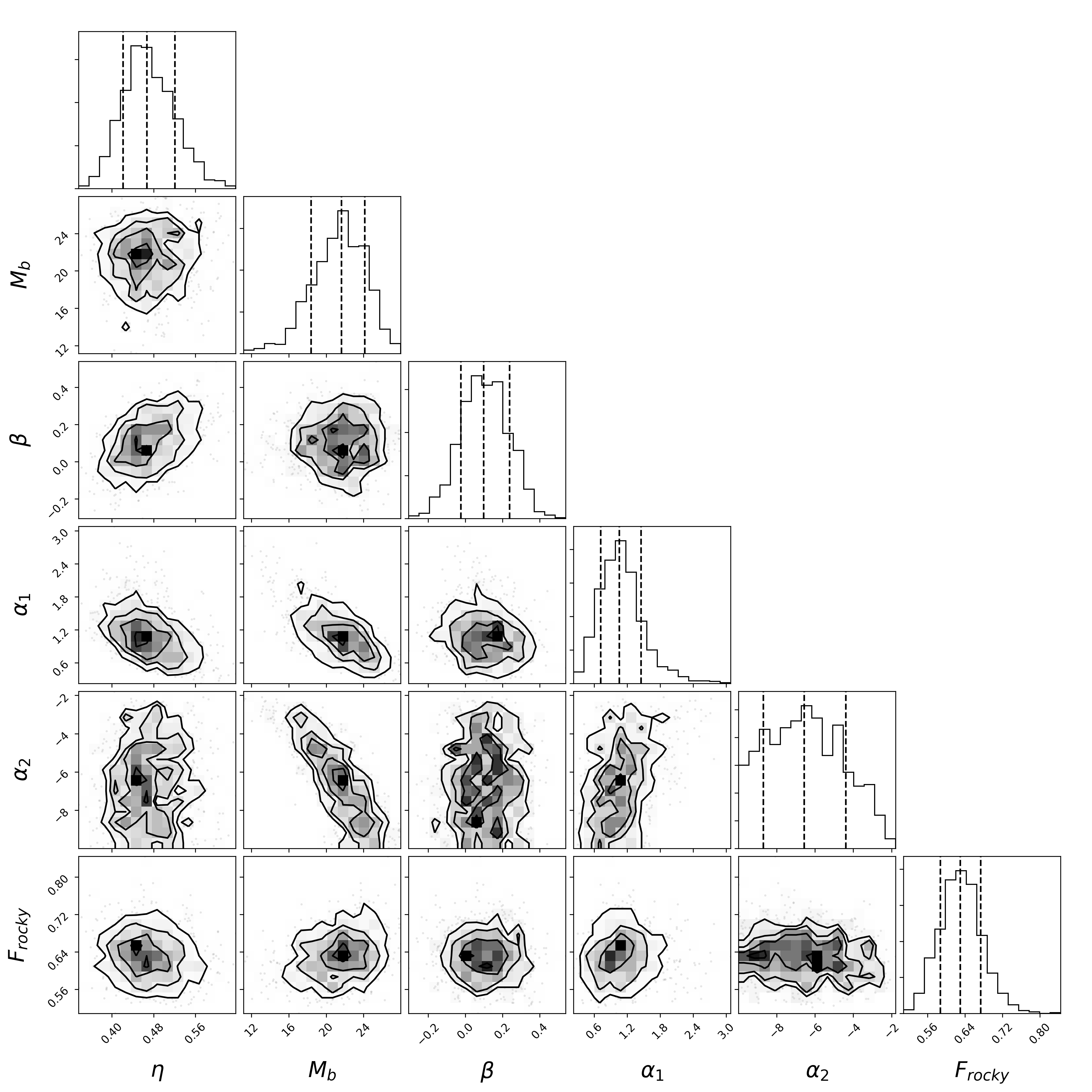}
    \caption{ABC posterior distributions for the combined RV + transit fit. The transit component includes correction for reliability. Each dotted line indicates the 15.9th, 50th, and 84.1th percentiles. Note that all histograms span a subset of our prior range (see Table (\ref{tab:reliability})).}
    \label{fig:combined}
\end{figure}

Our final combined results are shown in Figure (\ref{fig:combined}), giving $F_{\text{rocky}} = 0.63_{-0.04}^{+0.04}$, a slope of $\alpha_{1} = 1.3_{-0.3}^{+0.4}$ before a mass break at $M_{b} = 21.6_{-3.2}^{+2.4} M_{\oplus}$, and a slope of $\alpha_{2} = -6.6_{-2.1}^{+2.2}$ afterwards. The large uncertainties and flat posterior distribution of $\alpha_{2}$ are consistent with previous fits of \Kepler\ data to period-radius models, which have shown that the power law index for radius after the break is similarly unconstrained \citep{Mulders2018}.

\subsection{An Alternate Fit Removing a Low-Completeness Planet}
The RV dataset includes a single planet near $\sim2M_{\oplus}$, likely affecting the fit of the mass distribution due to its low completeness. As seen in the bias-corrected histogram of Figure (\ref{fig:rvhist}), this detection has an error bar contributing a number of planets spanning almost three orders of magnitude. 

As a bounding case, we re-fit our results using [3, 50] $M_{\oplus}$ as our mass range to exclude it. The RV-only mass break decreased to $M_{b} = 21.2_{-5.9}^{+6.2} M_{\oplus}$. This is closer to what we expect based on visually inspecting the histogram. However, $\eta$ dropped to $0.3_{-0.1}^{+0.1}$ planets per star, no longer within 1$\sigma$ of our transit-only results. 

Meanwhile, the joint RV + transit fit results remained essentially unchanged (see Table \ref{tab:altfits}), giving $M_{b} = 21.9_{-3.2}^{+2.3}$ $M_{\oplus}$ with [3, 50] $M_{\oplus}$ compared to $21.6_{-3.2}^{+2.4} M_{\oplus}$ with [2, 50] $M_{\oplus}$. The other parameters, including $F_{\text{rocky}}$, were also unchanged. This demonstrates the robustness of the joint fit, in that it is less sensitive to the inclusion or exclusion of the low-completeness planet.

\subsection{An Alternate Fit Using a Robust RV Sample}
We restrict our analysis even further to only include planets within [10, 50] $M_{\oplus}$. While this choice removes 4 (14\%) of the RV planets from the baseline sample, this mass range corresponds to a robust RV sample with modest completeness. As shown in Table \ref{tab:altfits}, our joint RV + transit fit recovered the same mass break at $M_{b} = 21.9_{-2.7}^{+2.4}$. All other parameters (with the exception of $\eta$, which is expected to be lower given the narrower mass range) are within 1$\sigma$ of our baseline $[2, 50] M_{\oplus}$ and alternate $[3, 50] M_{\oplus}$ results.

We note that this more reliable set of planets gives a power law index before the mass break of $\alpha_{1} = 0.6_{-1.3}^{+1.9}$. This is within $1\sigma$ of 0, which would correspond to a flat distribution in masses before $M_{b}$ rather than a turnover. However, the $1\sigma$ uncertainty is large, and $\alpha_{1}$ is more than $1\sigma$ from 0 when including the smaller planets. More data on the RV side and/or better characterization of the M-R relation for transiting planets in the overlap region could help clarify this behaviour. 

\renewcommand{\arraystretch}{1.5}
\begin{table}[h]
    \centering
    \caption{Joint Fit Results (With Reliability) Over Different Mass Ranges}\label{tab:altfits}
    \begin{tabular}{c |c c c| c}
    \hline\hline
        Parameter & $[2, 50] M_{\oplus}$ & $[3, 50] M_{\oplus}$ & $[10, 50] M_{\oplus}$ & Prior\\
        \hline
        $\eta$ & $0.47_{-0.05}^{+0.05}$ & $0.44_{-0.05}^{+0.04}$ & $0.32_{-0.04}^{+0.03}$ & $\mathcal{U}(0,3)$\\
        $M_{b}/M_{\oplus}$ & $21.6_{-3.2}^{+2.5}$ & $21.9_{-3.2}^{+2.3}$ & $21.9_{-2.7}^{+2.4}$ & $\mathcal{U}(X,50)$\footnote{Lower bound on the mass break prior was set to the start of the fitted mass range (2, 3, or 10 $M_{\oplus}$).}\\
        $\beta$ & $0.1_{-0.1}^{+0.1}$ & $0.1_{-0.1}^{+0.1}$ & $0.1_{-0.1}^{+0.1}$ & $\mathcal{U}(-5,5)$\\
        $\alpha_{1}$ & $1.1_{-0.3}^{+0.4}$ & $1.0_{-0.4}^{+0.5}$ & $0.6_{-1.3}^{+1.9}$ & $\mathcal{U}(-10,10)$\\
        $\alpha_{2}$ & $-6.6_{-2.1}^{+2.2}$ & $-6.7_{-2.1}^{+2.4}$ & $-7.4_{-1.7}^{+2.0}$ & $\mathcal{U}(-10,10)$\\
        $F_{\text{rocky}}$ & $0.63_{-0.04}^{+0.04}$ & $0.62_{-0.05}^{+0.05}$ & $0.64_{-0.05}^{+0.05}$ & $\mathcal{U}(0.6,1)$\\
    \end{tabular}
    \tablecomments{Central values and uncertainties are the median and 15.9th and 84.1th percentiles of the ABC posteriors.}
\end{table}

\section{Discussion}\label{sec:discussion}

\subsection{A Mass Break Comparison with Microlensing}

For the following comparison with microlensing studies, we report our results in terms of the planet-to-star mass ratio, $q$, which is a microlensing observable. Furthermore, our study is focused on FGK dwarfs with typical masses of $\sim1 M_{\odot}$, while microlensing studies focus on M dwarfs with typical masses of $\sim0.6 M_{\odot}$ \cite[e.g.][]{Suzuki2016}. \citet{Pascucci2018} found that the same break in the power law in $q$ can describe planets around G, K, and M stars regardless of spectral type, whereas the mass break is host star-dependent. This indicates that $q$ is a more readily comparable quantity between surveys targeting widely different stellar samples, and is also a more fundamental quantity in planet formation than planet mass.

\citet{Suzuki2016} and \citet{Pascucci2018} each noted a discrepancy between the breaks in power laws with $q$ as derived from \Kepler\ and microlensing. In particular, \citet{Pascucci2018} used the \citet{ChenKipping2017} M-R relation to convert \Kepler\ planet radii into masses, and found a planet-to-star mass ratio break of $q_{b} = (2.5\pm0.6)\times10^{-5}$ for planets orbiting G-type stars with $P < 100$ days and $R \sim1 - 6 R_{\oplus}$. This is at odds with results from microlensing that place the mass ratio break anywhere from $q_{b} \simeq 5.5\times10^{-5}$ \citep{Jung2019} to $q_{b} \simeq 1.65\times10^{-4}$ \citep{Suzuki2016}. Both power laws show increasing occurrence up to $q_{b}$, and decreasing occurrence towards larger $q$, meaning that this mass-ratio break marks a peak in planet occurrence rate. Given that microlensing probes planet orbits far beyond the \Kepler\ parameter space, these results would indicate that the most common planets beyond the snow line are $\sim2-8$ times more massive than those within the snow line.

Using the same M-R relation as \citet{Pascucci2018} applied to our simulated radii, we find $q_{b} = (2.5_{-0.3}^{+0.4})\times10^{-5}$ for \Kepler\ FGK planets with $25 < P < 200$ days and $R \sim1 - 8 R_{\oplus}$. This break is in strong agreement with \citet{Pascucci2018}. However, our mass ratio break increases to $q_{b} = (6.7_{-1.0}^{+0.8})\times10^{-5}$ when we adopt the \citet{Otegi2020} M-R relation and perform a combined RV + transit fit, which lies between the estimates from \citet{Jung2019} and \citet{Suzuki2016}. Notably, \citet{Suzuki2016} also provided Markov-Chain Monte Carlo results alongside their best-fit $q_{b} = 1.65\times10^{-4}$, finding $q_{b} = (6.7_{-0.18}^{+0.90})\times10^{-4}$, which is the same central value we find here. Overall, our joint RV/transit fit indicates that the most common planet beyond the snow line may be only $\sim1-3$ times more massive than within, though given the wide span in reported mass ratio breaks, more microlensing detections at low $q$ are likely required before we can determine the severity of any remaining discrepancy. 

We emphasize that this finding means that the \citet{Otegi2020} M-R relation allows us to bridge the mass break gap between the transit population and both RV and microlensing populations. We consider this evidence that it is a more accurate representation of the observed planet population.

\subsection{An Estimate of the Fraction of Rocky Planets Between 5 and 25 $M_{\oplus}$}

By using the \citet{Otegi2020} M-R relation, we are also able to place an estimate on the average fraction of rocky planets between $5$ and $25$ $M_{\oplus}$, finding $F_{\text{rocky}} = 0.63_{-0.04}^{+0.04}$. Encouragingly, 18 of the planets in the catalogue used by \citet{Otegi2020} that lie in this mass range are considered rocky, and 7 are considered volatile-rich. This corresponds to an observed rocky fraction of $F_{\text{rocky}} \sim 0.7$. Given that this catalogue suffers from observational biases since it is easier to detect more massive planets at a given radius, this should be considered an upper limit on the true fraction. Our inferred $F_{\text{rocky}}$ lies below this upper limit.

\citet{WolfgangLaughlin2012} performed a similar investigation regarding the apparent discrepancy in the numbers of planets per star inferred from the HARPS and \Kepler\ surveys for $M \lesssim 17 M_{\oplus}$ ($R \lesssim 4 R_{\oplus}$) and $P < 50$ days. They were able to find a consistent planet frequency only by adopting a mixture of rocky and gaseous M-R relations for planets between 1 and 17 $M_{\oplus}$. More recently, \citet{NeilRogers2020} used the \Kepler\ population to constrain a joint mass-radius-period distribution of exoplanets and found that mixture models that differentiated between planets with gaseous envelopes, evaporated cores, and intrinsically rocky planets were preferred over the baseline model with a single population of planets with compositions ranging from rocky to gaseous. Overall, our results provide strong evidence for the existence of distinct planet populations across transition regions in mass or radius. We suggest that future M-R relations account for this in order to more accurately represent the observed population.

\section{Caveats and Limitations}\label{sec:caveats}

Here, we summarize important caveats, and the main assumptions and design choices made in our study. Each of these can motivate future improvements to our occurrence rate model and methodology.

\textit{Reliability:} We were unable to take into account reliability of the RV population against astrophysical or noise/systematic false positives because its false positive rate was not characterized.  We do not believe this has a significant impact on our analysis because we expect the HARPS + CORALIE survey has high reliability, and we explored the more robust $> 10 M_{\oplus}$ regime, where we assume the false positive rate is lower, and still found results consistent with our baseline analysis. Nevertheless, we would like to be able to directly take into account reliability for any given survey in the future, and encourage planet searches to characterize their false positive rates.

\textit{Detection Efficiency:} For both surveys, we adopted an average completeness when calculating the probability of detecting a planet with a given set of properties. As discussed in \citet{Hsu2019} and \citet{KunimotoMatthews2020}, a star-dependent detection probability is likely more accurate given how rapidly it changes with signal-to-noise ratio at low completeness. Furthermore, using target-dependent window functions has been shown to reduce occurrence rates for planets at long orbital periods ($\gtrsim 1$ year) because they better estimate detection probability when there are few transits \citep{Hsu2019}. Given that we are only looking at planets with $P < 200$ days, and we did not reach the limits of \Kepler\ detection sensitivity with our chosen period and mass range, we do not expect our adoption of average detection efficiencies to have a significant effect on our results.

\textit{Source of RV Data:} Our RV results are reliant on a single study from \citet{Mayor2011} with only 29 planets in our domain of analysis. This paper has not yet been accepted, though it has been cited in several previous occurrence rate studies \cite[e.g.][]{Wright2012, Santerne2016, Fernandes2019}. We chose this survey because it is one of the only RV planet discovery surveys with well-characterized completeness beyond simple detection limits. It also has sensitivity down to small planets around FGK stars. We would like to extend our analysis with a larger dataset, such as the ongoing work by the California Planet Search (CPS)\footnote{https://exoplanets.caltech.edu/cps/}, and incorporate multiple RV surveys in the same joint fit to fully take advantage of available constraints.

\textit{Planet Population Model:} We assumed a planet distribution function in the form of a joint power law in orbital period and planet mass, with a single break in mass (Eqn. \ref{eqn:powerLaw}). We also assumed that these power laws are independent. However, there is much evidence that exoplanet populations are significantly more complex \citep{Fulton2017, Mulders2019, Pascucci2019}. We could also explore models where the different sub-populations of planets (e.g. rocky and volatile-rich) follow different mass-period distributions.

\textit{Rocky Planet Fraction Model:} Our model for the fraction of rocky planets, $F_{\text{rocky}}$, was only a simple constant over the $M  = 5 - 25 M_{\oplus}$ overlap range quoted in \citet{Otegi2020}. $F_{\text{rocky}}$ could have a more complicated behaviour that depends on mass within the transition. With the current TESS mission adding more masses for planets with $R < 4 R_{\oplus}$ as one of its primary science goals \citep{Ricker2014}, we anticipate being able to better characterize its features in the near future.

\section{Conclusions and Future Work}

In this paper, we present the first joint fit analysis that combines independent ground RV survey and \Kepler\ transit survey data, both with well-characterized completeness. We assumed a planet distribution function described by a simple power law in period and broken power law in mass, and fit this model to the results of each survey using a forward modeling approach with ABC. 

In summary, our main findings are as follows:

\begin{itemize}
    \item A single planet population model for planets across $M = [2, 50] M_{\oplus}$ and $P = [25, 200]$ days can be consistent with both RV and transit surveys when we adopt the M-R relation of \citet{Otegi2020}, which allows for overlapping sub-populations of rocky and volatile-rich planets to exist between the masses of $M = 5 - 25 M_{\oplus}$. Using an M-R relation without such an overlap, the mass break found for the transiting population is significantly lower than what is found for RV.
    \item Taking advantage of constraints from both types of surveys simultaneously, our joint fit finds a mass break at $M_{b} = 21.6_{-3.2}^{+2.5} M_{\oplus}$. The joint fit also finds that the transit and RV populations are most consistent when a fraction of $F_{\text{rocky}} = 0.63_{-0.04}^{+0.04}$ planets belong to the rocky population across $5 - 25 M_{\oplus}$, assuming a simple, constant fraction of rocky planets in this overlap region.
    \item The joint-fit mass break after conversion to a planet-to-star mass ratio (assuming a typical star mass of $1 M_{\odot}$) is consistent with estimates from microlensing, while again the non-overlapping M-R relation finds a mass ratio break significantly lower. Thus, the \citet{Otegi2020} relation allows us to resolve apparent discrepancies with both RV and microlensing populations. 
\end{itemize}

These findings contribute to a growing body of evidence for the existence of distinct types of planets overlapping in mass and radius in the observed planet population, which should be accounted for by future M-R relations. Our statistical analysis was only possible thanks to the availability of both RV and transit surveys that have provided estimates of their completeness. We encourage future teams to design their surveys to be amenable to statistical population studies.

The ability to perform simultaneous fits between various RV and transit surveys will become increasingly more important as we enter the next era of exoplanet discovery. For instance, survey yields from the TESS mission could be fit alongside \Kepler\ to more deeply explore the planet distribution function at short orbits, should the completeness of TESS be sufficiently characterized. The next generation of RV spectrographs such as EXPRES and ESPRESSO are also expected to be able to detect small, Earth-size planets around M dwarf stars for the first time. Given that \Kepler\ observed only a few thousand M dwarfs, being able to take advantage of the additional constraints provided by RV could significantly improve understanding of exoplanet occurrence rates, especially the occurrence rate of potentially habitable planets around these stars. 

Furthermore, while only an RV + transit example was presented here, our framework can be extended to include other detection methods in the future. In particular, the Roman Space Telescope microlensing survey will be complementary to the \Kepler\ survey in finding hundreds of small planets in orbits of up to thousands of days \citep{Penny2019}, while \textit{Gaia} is expected to find thousands down to the masses of Neptune using astrometry \citep{Ranalli2018}. A joint fit analysis incorporating all of these types of surveys could facilitate the most comprehensive statistical census of exoplanets to date.

\acknowledgments
We are grateful for the referee's comments and constructive input, which substantially improved the manuscript. We thank Leslie Rogers for helpful comments and insight. We also thank the HARPS+CORALIE and \Kepler\ teams for providing the data that make these studies possible.

\software{\texttt{cosmoabc} \citep{Ishida2015}, \texttt{epos} \citep{Mulders2018}}

\bibliography{references}

\end{document}